# Conformal Antenna Array for Millimeter-Wave Communications: Performance Evaluation

V. Semkin[1], A. Bisognin[2,3], M. Kyrö[4], V-M. Kolmonen[5], C. Luxey[2,6], F. Ferrero[7], F. Devillers[8], and A.V. Räisänen[1]

[1] Aalto University, P.O. Box 13000 FI-00076 AALTO, Finland.

[2] EpoC, Université Nice-Sophia Antipolis, 06560 Valbonne, France.

[3] STMicroelectronics, 38926 Crolles, France.

[4] Finnish Patent and Registration Office, P.O. Box 1140, FI-00101 Helsinki, Finland.

[5] Philips Medical Systems MR Finland.

[6] Institut Universitaire de France, 75005 Paris, France.

[7] LEAT-CREMANT CNRS, Université Nice-Sophia Antipolis, 06560 Valbonne, France.

[8] Orange Labs-CREMANT, 06320 La Turbie, France.

*In this paper, we study the influence of the radius of a cylindrical supporting structure on radiation properties of a conformal millimeter-wave antenna array. Bent antenna array structures on cylindrical surfaces may have important applications in future mobile devices. Small radii may be needed if the antenna is printed on the edges of mobile devices and in items which human beings are wearing, such as wrist watches, bracelets and rings. The antenna under study consists of four linear series-fed arrays of four patch elements and is operating at 58.8 GHz with linear polarization. The antenna array is fabricated on polytetrafluoroethylene substrate with thickness of 127 µm due to its good plasticity properties and low losses. Results for both planar and conformal antenna arrays show rather good agreement between simulation and measurements. The results show that conformal antenna structures allow achieving large angular coverage and may allow beam-steering implementations if switches are used to select between different arrays around a cylindrical supporting structure.*

Corresponding author: V. Semkin; email: vasilii.semkin@aalto.fi

# I. INTRODUCTION

During the last decades conformal antennas have been studied very actively because they can offer operational advantages over planar structures. Conformal antennas are more common at lower microwave frequencies. One of the main benefits is that conformal antennas replicate the form of an object on which they are placed. Using this advantage allows antenna integration in different applications such as satellite communication, communication with airplanes and mobile communications. Using the surfaces of airplane wings or hulls provides better aerodynamic characteristics because these antennas are flush-mounted and do not protrude. Furthermore, they do not require specially developed radomes [1].

Conformal antennas can also be used in Wireless Local Area Networks (WLAN) to achieve high speed communications. The main benefit from using conformal structures can be obtained if the surface on which the antenna is placed can dynamically change its shape [2]. In this case, the radiation properties can be reconfigured according to the changes in the communication channel or in the system requirements, i.e. broad beam radiation patterns can be achieved and the realized gain values can be controlled with the shape of the antenna structure.

Nowadays, an increasing interest in high data rate transmissions for multimedia applications over short distances and requirements drives recent developments in millimeter wave frequencies. Allocation of wireless networks operating at mm-wave frequencies is one of the promising solutions to support communications with data rates of several Gbit/s. At those frequencies, very wide bandwidths are available and such bands have already been allocated around the world for high speed wireless communications [3]. The 60 GHz frequency with 5 mm free-space wavelength makes the antenna structure physically small. The bandwidth of 9 GHz available at 60 GHz range [4] provides a high data rates and a possibility to avoid co-channel interference. Although, free-space losses at this frequency are rather high, utilizing antenna array with a high gain can compensate large free space losses. For indoor radio links, the human blockage can also have high impact on the radio channel. Antennas with high gain and beam steering capabilities are needed to overcome this problem.

To date, the antenna research at millimeter-wave frequencies has mostly concentrated on planar configurations. A common solution for 360˚ area coverage is to use an array integrated on the four faces of a cube with two antennas realized on each face, such as in [5]. A different approach is presented in [6] where multi-faceted phased arrays are investigated. Other approaches are studied in [7] and [8], where conformal microstrip antenna arrays are presented for continuous conformal surfaces at 35 and 32 GHz. Conformal substrate with integrated waveguides has been developed at 35 GHz in [9]. The latter approach has been selected as a framework for experimental study of patch antenna arrays operating at 60 GHz.

In this work, we concentrate on cylindrical array configurations, where the ultimate target is a wide scan angle implementation. The investigation of an antenna array that achieves high gain and covers a large angular range with symmetric beams has been done. In this study, the antenna is placed on a cylindrical-shape object and radiation patterns are measured. One of the possible implementations of the structure mentioned above is shown in Fig. 1. Several conformal antenna arrays can be integrated into a bracelet which can be wearable on the wrist. This solution can be beneficial in different devices, such as smart watches or as a transmitter during an ultrasonic inspection of the patient.

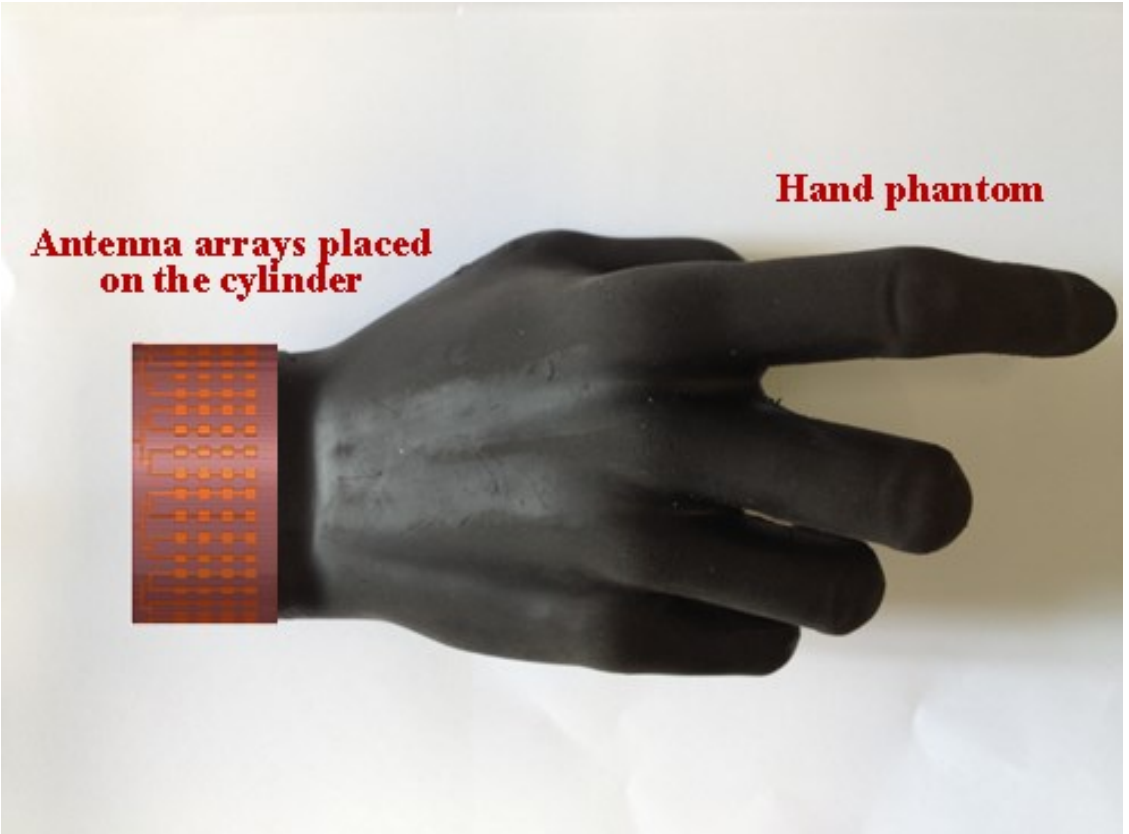

Fig. 1. Possible antenna integration in a wrist bracelet.

The goal of this work is to evaluate the performance, in particular gain and coverage, of series fed arrays working at 60 GHz frequency band, which are bent around a cylindrical surface with different radii. Radiation characteristics of the proposed antenna structures are analyzed. Configuration with switches selecting between different arrays is beyond the scope of this study but is a goal in future studies.

Section II describes the antenna design and 4x4 series-fed antenna array structure. In Section III, we present simulation and measurement results for the planar and different conformal cases. Finally, Section IV provides discussion and conclusions.

## II. DESIGN OF THE ANTENNA STRUCTURE

Microstrip antenna arrays are the most suitable antennas in a conformal installation because of their low profile structure, simple geometry and relatively inexpensive manufacturing. Printed circuit technology can be used to integrate these antennas with millimeter wave circuits [8], for example, with the smart watches. In addition, according to the IEEE 802.11ad standard, single carrier modulation scheme MCS-6 (QPSK) can deliver 1.5 Gbit/s which is enough for high definition (HD) video transmission. The minimum required power level for the MCS-6 is -63 dBm [10].

We aim at an antenna array that has a resonant frequency of 58.8 GHz. Free space path loss at 58.8 GHz can be derived from the Friis equation and calculated by:

$$L_{FS} = 32.45 + 20\log(R) + 20\log(f) \qquad (1)$$

where $L_{FS}$ is free space loss, $R$ is distance in kilometers and $f$ is the resonant frequency in MHz [11]. The intended applications of the developed antenna array are indoor scenarious, such as conference rooms, concert and exhibition halls, and hospitals. The maximum distance for the radio link will be approximately 20 m. At 60 GHz the free space loss at this distance is 93.85 dB. The maximum equivavalent isotropic radiated power (EIRP) is 40 dBm, and the maximum transmit power, which is reduced for safety reasons, is 10 dBm [12]. Taking into account the regulations mentioned above and losses it is preferable to use an antenna with a high gain value. The required antenna gains can be estimated from the Friis equation:

$$\frac{P_r}{P_t} = G_t G_r \frac{1}{L_{FS}} \tag{2}$$

where $P_r$ is received power, $P_t$ – transmit power, $G_t$ – transmitter gain, $G_r$ – receiver gain, and $L_{FS}$ is a free space loss [13]. In order to get -63 dBm received power level with free space loss of 93.85 dB and 10 dBm transmit power, the antennas (transmitter and receiver) should have gain values more than 10.4 dB (for the case $G_t=G_r$), which is calculated from Equation (2). This value is set as a requirement for the design of the microstrip patch antenna array. Simultaneously, obtaining large azimuth coverage by bending the antenna will be beneficial in the crowded premises where the line-of-sight (LOS) radio link can be blocked by the human beings so that the strongest reflected path can be used. In addition, switching between several antenna arrays, as shown in the configuration in Fig.1, can improve the received power level. Either a conformal or planar antenna can be used as a transmitter (Tx) and a conformal antenna integrated into smart watches or rings can be used as a receiver (Rx).

### A) Substrate

For conformal antenna realization on the cylindrical surface substrate material should have good plasticity properties. Several flexible substrate materials are available for manufacturing such as liquid crystal polymer (LCP), polyethylene terephthalate (PET), polyethylene naphthalate (PEN), and Fiber Teflon [14, 15]. It is obvious that the radius of the supporting cylinder is a significant design parameter both from the antenna performance for the mechanical feasibility point of view. A polytetrafluoroethylene (PTFE) material from Taconic (TLY-5) was chosen for manufacturing the antenna array due to its low losses and good flexibility properties. The relative permittivity of the substrate is $\varepsilon_r$=2.2 and the dielectric loss tangent is tan$\delta$=0.0009 (material properties are specified at 10 GHz). Substrate thickness is 0.127 mm, and the antenna structure has a ground plane of the same size as the substrate. Metallization thickness is 0.017 mm.

### B) Series-Fed Antenna Array

First, the dimensions of the patch elements, such as width and length, were calculated using conventional methods [16] and after that the size of the elements in the array was optimized to get the resonant frequency at 58.8 GHz. The layout of the antenna structure is presented in Fig. 2. The antenna consists of 4x4 patch elements placed on top of the PTFE material substrate. The overall size of the antenna array is 25x20 mm$^2$. Based on the simulation results, single element dimensions are determined as 1.65x1.7 mm$^2$. Microstrip signal and feed line widths are 0.3 mm and 0.12 mm, respectively. This prototype is bent around H-plane axis of the patches (XZ plane). The manufactured antenna array is shown in Fig. 3. Simulated antenna performance is presented and compared with measured performance in Section 3.

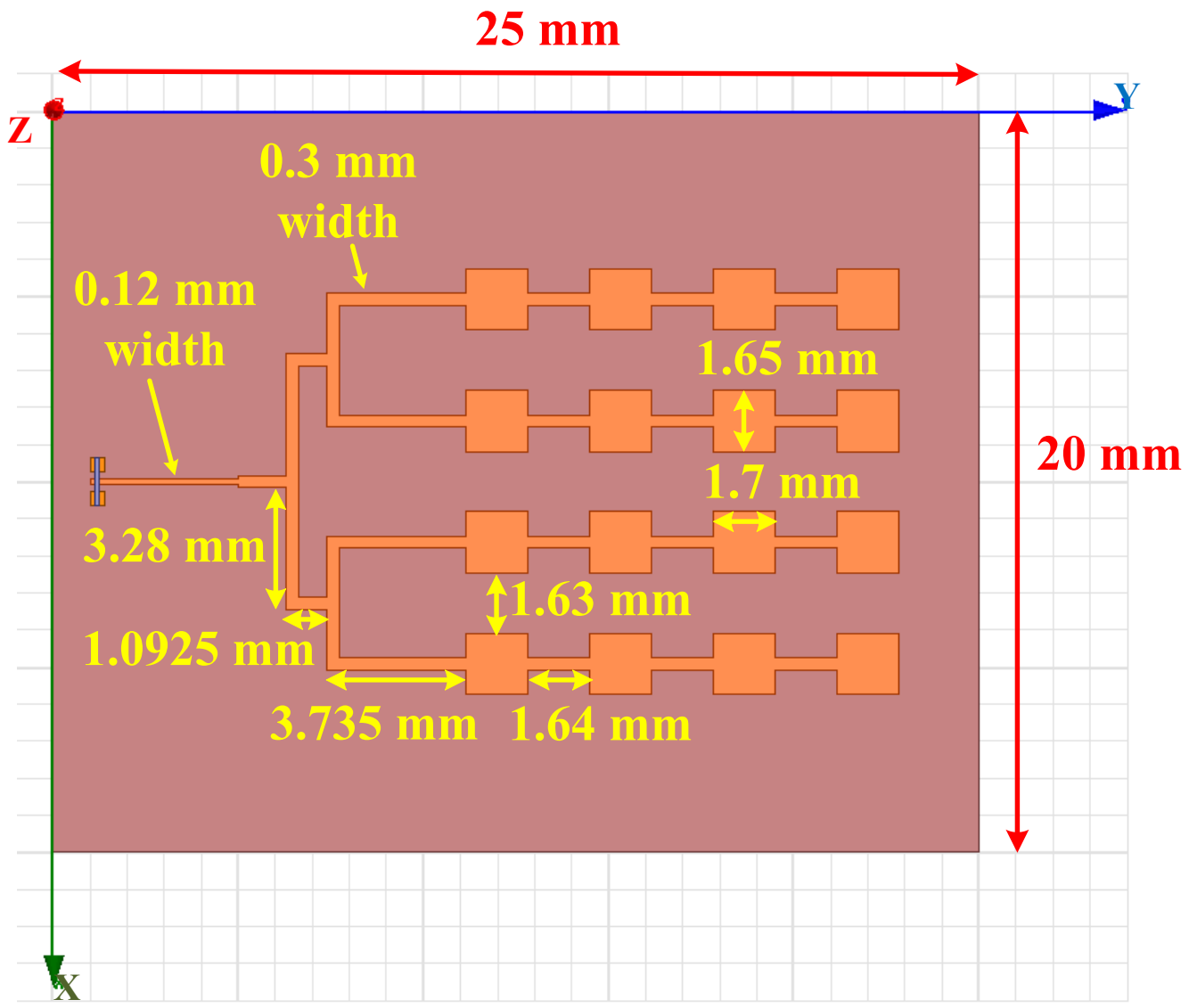

Fig. 2. Top-view of the series fed array.

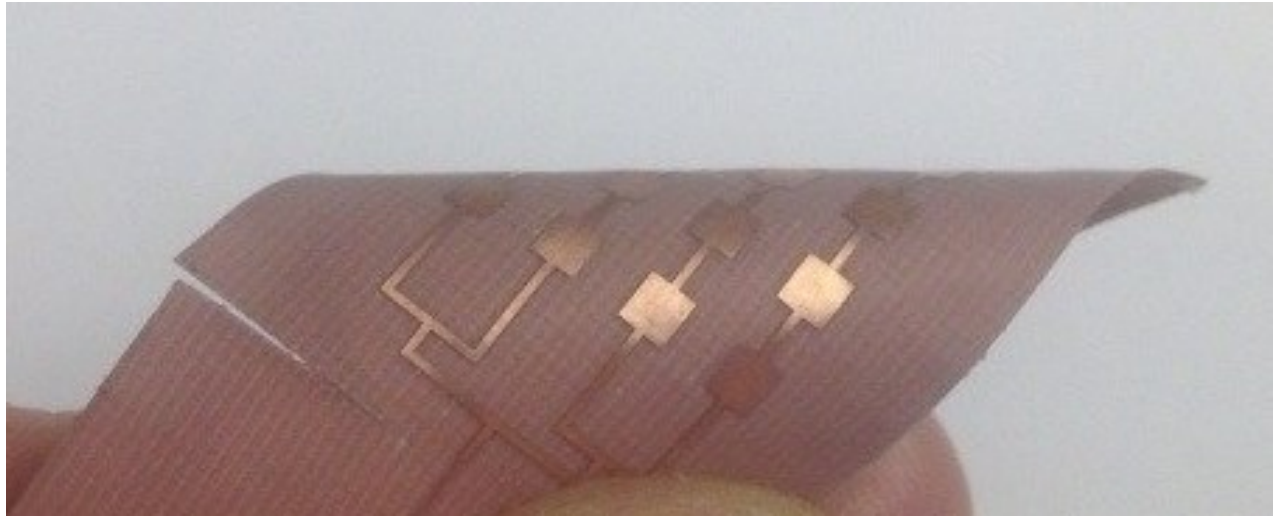

Fig. 3. Antenna array structure manufactured on PTFE material (Taconic TLY-5).

## III. MEASUREMENT AND SIMULATION RESULTS

The measurement system used in this work and presented in [17] allows measuring quasi three dimensional sphere over the antenna under test (AUT). Two arms are used to rotate a horn antenna over the foam supporting structure with the probe-fed antenna on it (Fig. 4). Two laser pointers were used to position the AUT in the middle of the sphere. This measurement system allows measuring particularly 84% of the total sphere of the radiation pattern. A detailed description of the measurement system and the methods used in the post processing can be found in [17]. The feeding part of the series-fed antenna array has been optimized to cope with a Ground-Signal-Ground (GSG) probe feeding approach [17]. The distance between the AUT and the horn antenna is 20 cm. The largest dimension D of the array aperture has to be smaller than 21 mm to comply with the far-field criterion ($\frac{2D^2}{\lambda}$) of the measurement system. The frequency range from 50 GHz to 67 GHz of the measurement equipment was used in the measurements.

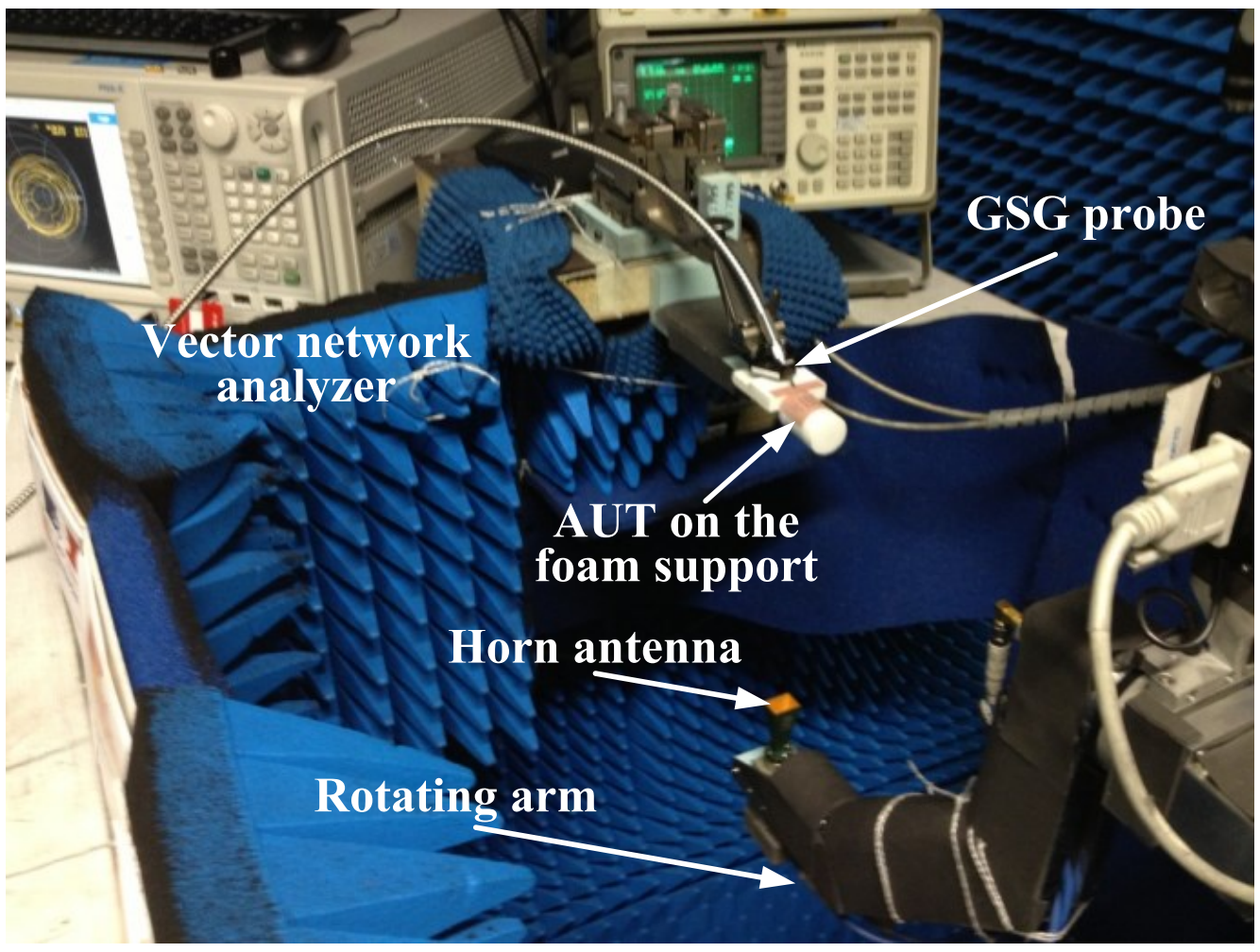

Fig. 4. A photo of the measurement setup.

The antenna array structures were simulated using 3D full-wave electromagnetic field software Ansoft HFSSTM. Supporting material for the antenna arrays is styrofoam with relative permittivity $\varepsilon_r$ = 1.03. Due to the reason that the antenna has a ground plane; other supporting structures do not have a significant effect to the radiation performance. The series fed antenna array is matched at 58.8 GHz. Simulations and measurements of these antenna structures were done for four different cases (Fig. 5(a):

- planar antenna structure,
- bent on a cylinder with the radius R = 12 mm,
- bent on a cylinder with the radius R = 8 mm,
- bent on a cylinder with the radius R = 6 mm.

The relative permittivity of the supporting foam material is very close to that of air, therefore, the impact of the styrofoam support is negligible. An example of the simulated structure is shown in Fig. 5(b). Supporting foam structure is not shown for simplicity.

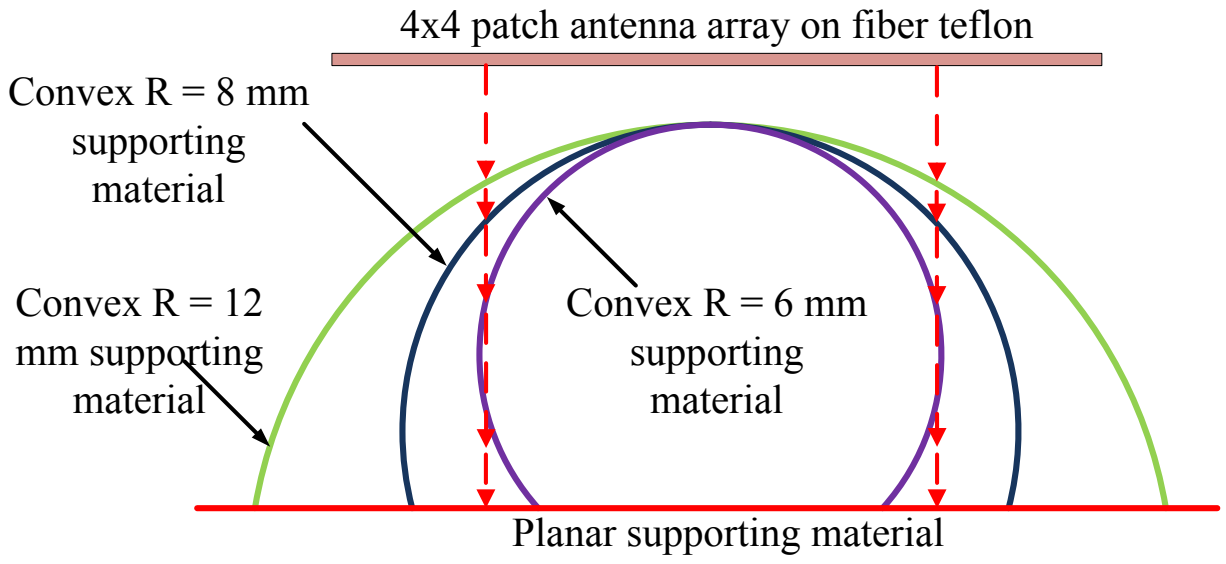

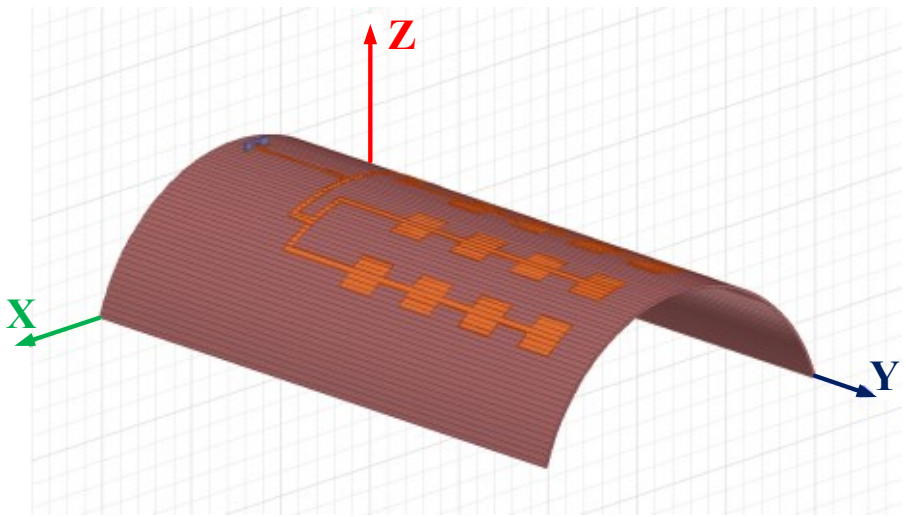

(a) (b)

Fig. 5. (a) Sketch of the planar an convex antenna array structures, (b) antenna array model in HFSS and the used coordinate system (convex state on the cylinder with radius $R$ = 6 mm).

The measurement setup is presented in Fig. 6. A 400 μm (GSG) pad is used to feed the antenna arrays. To ensure GSG-probe feeding for the conformal cases, cuts in the antenna substrate were made so that the probing part of the antenna remains planar and the antenna array is bent around the cylindrical surface (Fig. 6).

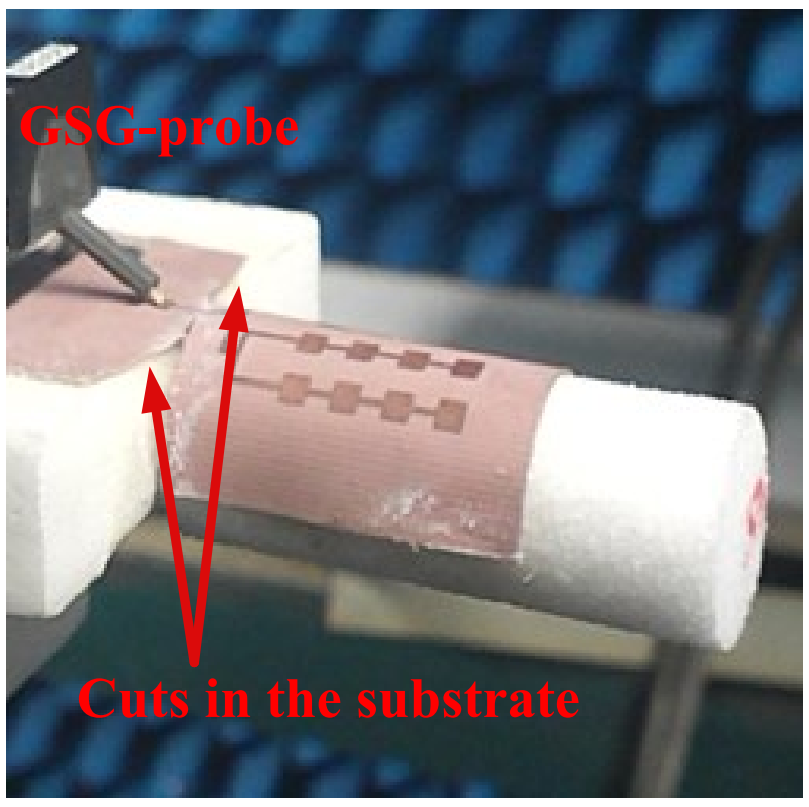

Fig. 6. Photograph of the supporting foam and the probe-fed 4x4 patch antenna array, convex structure $R = 6$ mm.

Next, the simulation and measurement for the series-fed antenna array are analyzed. Realized gain values to the broadside direction of the series-fed antenna array versus frequencies are presented in Fig. 7. Receiving horn antenna was set above the antenna array and the maximum realized gains versus frequency were measured along the X-axis. Linear polarization over the Y-axis direction is obtained. The maximum measured realized gain values for all 4 cases are summarized in Table 1 and compared with the simulated ones. It can be observed that with the decreasing flexing radius of the structure the measured gain value reduces from 17.2 dB to 11.4 dB, and corresponding widening of the beam width of the main lobe can be observed. Several antenna samples were fabricated and measured. The difference between the measured realized gain values of different samples is insignificant.

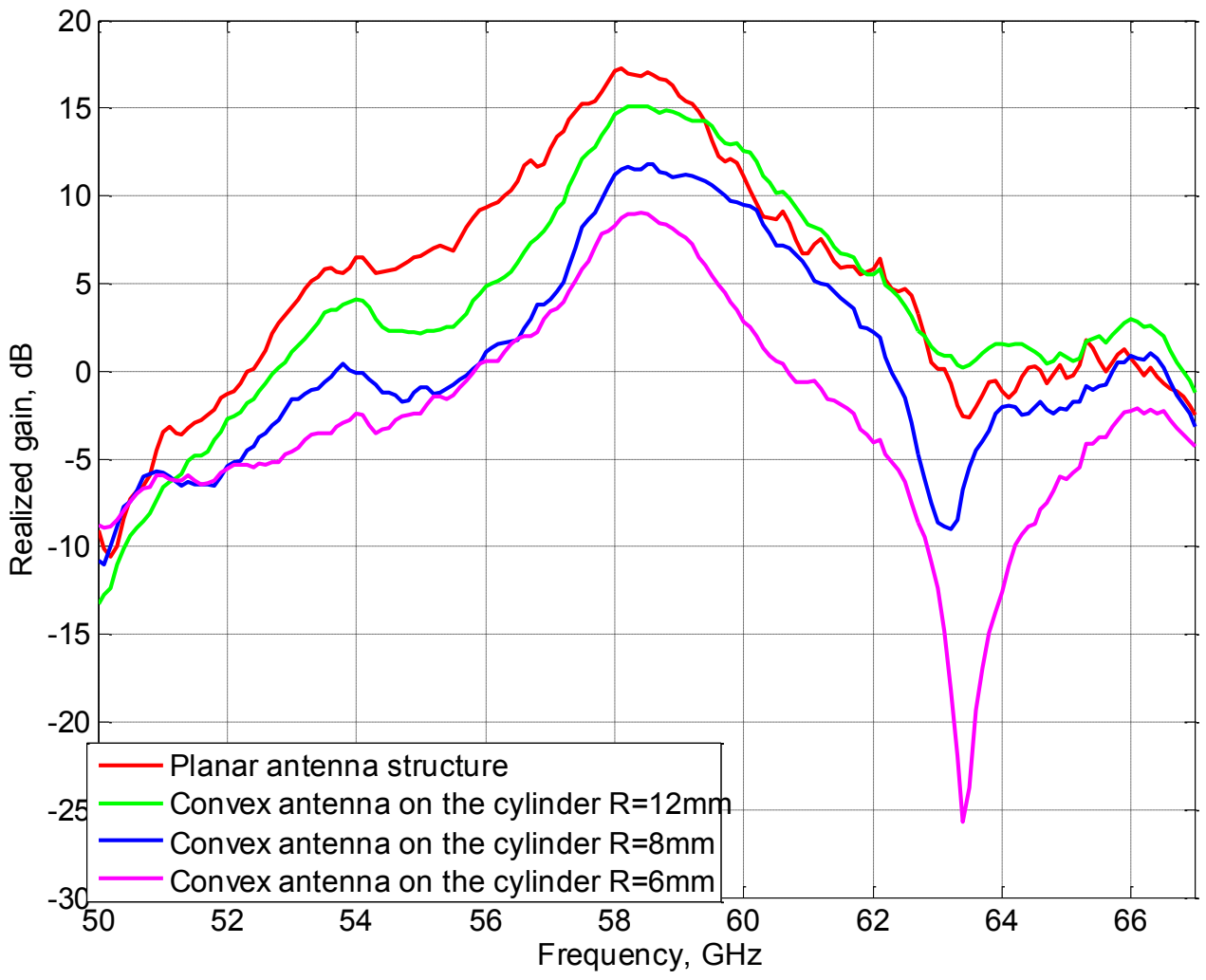

Fig. 7. Measured realized gain values for planar and conformal cases for the antenna array.

A comparison between the simulated and measured realized gain values in the XZ-plane is presented in Fig. 8. The measured radiation patterns correspond well to the simulated ones. The disagreement may be explained by losses in the substrate material because there is no information available about the losses at 60 GHz for the Taconic TLY-5. Also ohmic losses within the distribution network that are underestimated could degrade the measurement results. A comparison of the simulated and the measured 3D realized gain radiation patterns is given in Fig. 9.

Table 1. Comparison of simulated and measured gain and HPBW in XZ-plane.

| **4x4 series fed antenna array** | **Simulated gain (@ 58.8 GHz)** | **Measured gain (@ 58.5 GHz)** | **Simulated HPBW** | **Measured HPBW** |
|---|---|---|---|---|
| **Planar** | 19.0 dB | 17.2 dB | 20° | 20° |
| **Convex, R=12mm** | 16.9 dB | 15.2 dB | 23° | 23° |
| **Convex, R=8mm** | 14.6 dB | 12.3 dB | 62° | 72° |
| **Convex, R=6mm** | 12.4 dB | 11.4 dB | 82° | 79° |

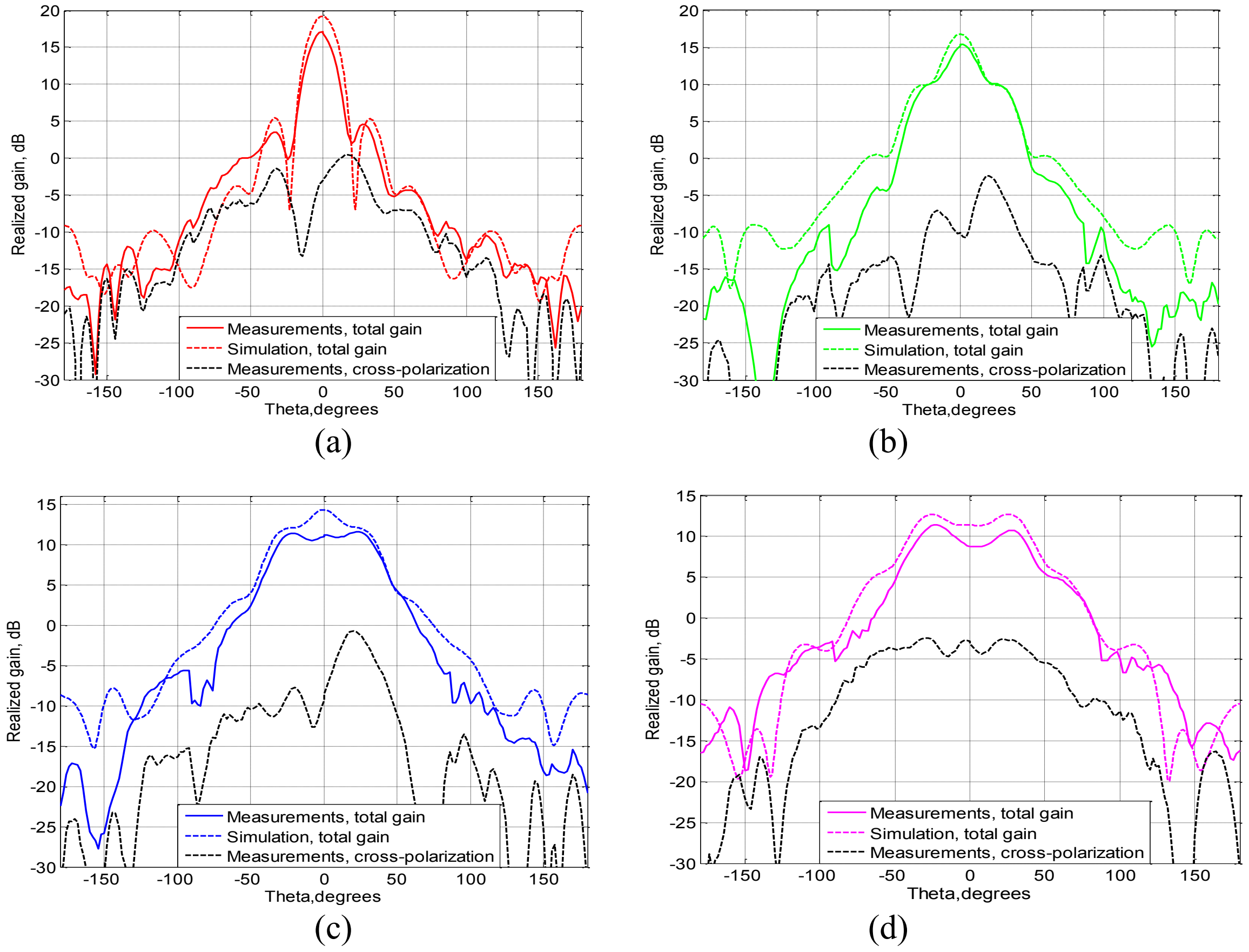

Fig. 8. Simulated and measured total and cross-polarization realized gain radiation patterns in XZ-plane (a) planar structure, (b) bent on the cylinder with radius $R$ = 12 mm, (c) bent on the cylinder with radius $R$ = 8 mm, (d) bent on the cylinder with radius $R$ = 6 mm.

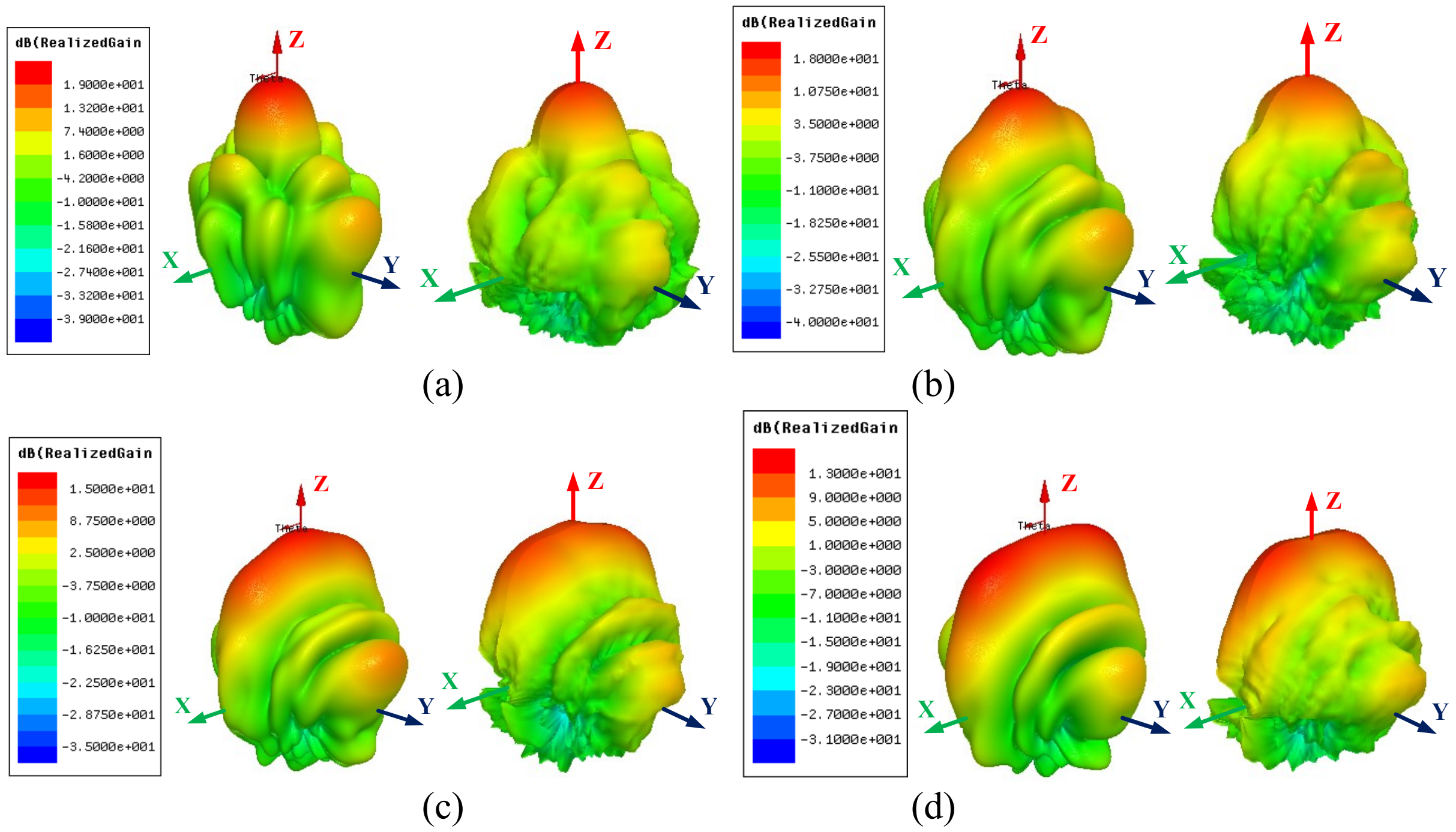

Fig. 9. Simulated (left) and measured (right) 3D total realized gain radiation pattern (a) planar structure, (b) bent on the cylinder with radius $R$ = 12 mm, (c) bent on the cylinder with radius $R$ = 8 mm, (d) bent on the cylinder with radius $R$ = 6 mm.

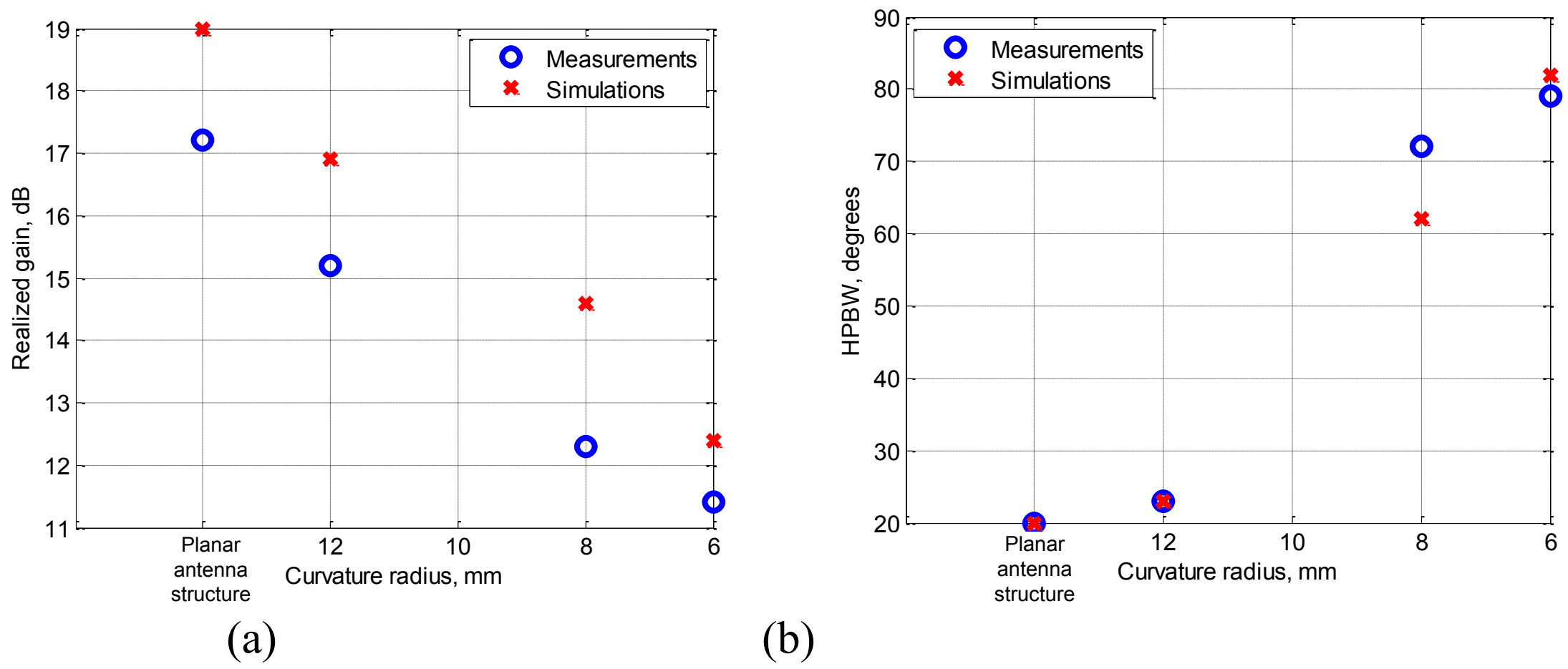

Fig. 10. Dependence of (a) realized gain versus curvature radii and (b) HPBW from curvature radii (blue markers denote to measurement results, red markers denote to the simulation results).

In Fig. 10 (a, b) the measured and simulated dependence of realized gain and half-power beamwidth (HPBW) versus different radii are presented. With the decreasing cylinder radius the realized gain values also decrease. For the planar case the realized gain is 17.2 dB and for the conformal cases the realized gain nearly linear increases with the radius of the supporting structure. Therefore, one results give a possibility to estimate the realized gain from the antenna geometry. Corresponding, enlargement in the HPBW is also visible. For the planar case HPBW is 20˚ and for the smallest radius (R = 6 mm) HPBW is almost four times larger.

## IV. DISCUSSION AND CONCLUSION

Simulations and measurements of a series fed antenna array at 58.5 GHz were performed for different curvature radii of the supporting structure. It is verified that the realized total gain of the antenna array decreases with the diminution of the radius of the supporting structure and some widening of the main beamwidth is observed. For the planar case of the series-fed antenna array realized gain value is 17.2 dB and HPBW is 20°, and for the bent antenna on the cylinder with radius 6 mm the realized gain is 11.4 dB and HPBW is 82°. The conformal antenna array bent on the cylinder with radius R = 6 mm has 1 dB higher gain than it was set in the requirements for 1.5 Gbit/s communications.

The antenna design may be improved after verification the losses of the material at 60 GHz. By performing these simulations and measurements we have shown that the optimal beamwidth of the conformal antenna array for the particular scenario can be defined. Beamwidth widening property can be used for increasing the antenna coverage area, i.e., several antenna arrays placed on the cylinder structure will allow covering the 360° area. Omnidirectional radiation pattern with high gain values could be achieved. Utilization of the conformal antenna arrays with these parameters in conjunction with a switching network allows implementation of beam steering technology. The main beam can be pointed in another direction if the line-of-sight is blocked and the strongest multipath component can be used. This can help to overcome losses due to a possible blockage in an indoor scenario.

## ACKNOWLEDGEMENT

This work was supported in part by the Finnish Funding Agency for Technology and Innovation (Tekes) under the BEAMS project. The authors would like to thank all the project partners. This work was also supported in part by the Academy of Finland under the DYNAMITE project. V. Semkin would like to thank the Graduate School of Electronics, Telecommunications and Automation (GETA) of Finland, Aalto ELEC doctoral school, and Foundation of Nokia Corporation for financial support.